\documentclass{vgtc}                          

\ifpdf
  \pdfoutput=1\relax                   
  \usepackage{graphicx}                
  \DeclareGraphicsExtensions{.pdf,.png,.jpg,.jpeg} 
\else
  \usepackage{graphicx}                
  \DeclareGraphicsExtensions{.eps}     
\fi%

\graphicspath{{figures/}{pictures/}{images/}{./}} 

\usepackage{microtype}                 
\PassOptionsToPackage{warn}{textcomp}  
\usepackage{textcomp}                  
\usepackage{mathptmx}                  
\usepackage{times}                     
\usepackage{cite}                      
\usepackage{tabu}                      
\usepackage{booktabs}                  
\usepackage{url}

\onlineid{1032}

\vgtccategory{Technical}

\newcommand{\thename}{\textit{EQSA}}

\title{\thename{}: Earthquake Situational Analytics from Social Media
}

\author{Huyen N. Nguyen\thanks{e-mail: huyen.nguyen@ttu.edu}\\ %
        \scriptsize Department of Computer Science, Texas Tech University %
\and Tommy Dang\thanks{e-mail: tommy.dang@ttu.edu}\\ %
     \scriptsize Department of Computer Science, Texas Tech University %
}

\teaser{
  \includegraphics[width=\linewidth]{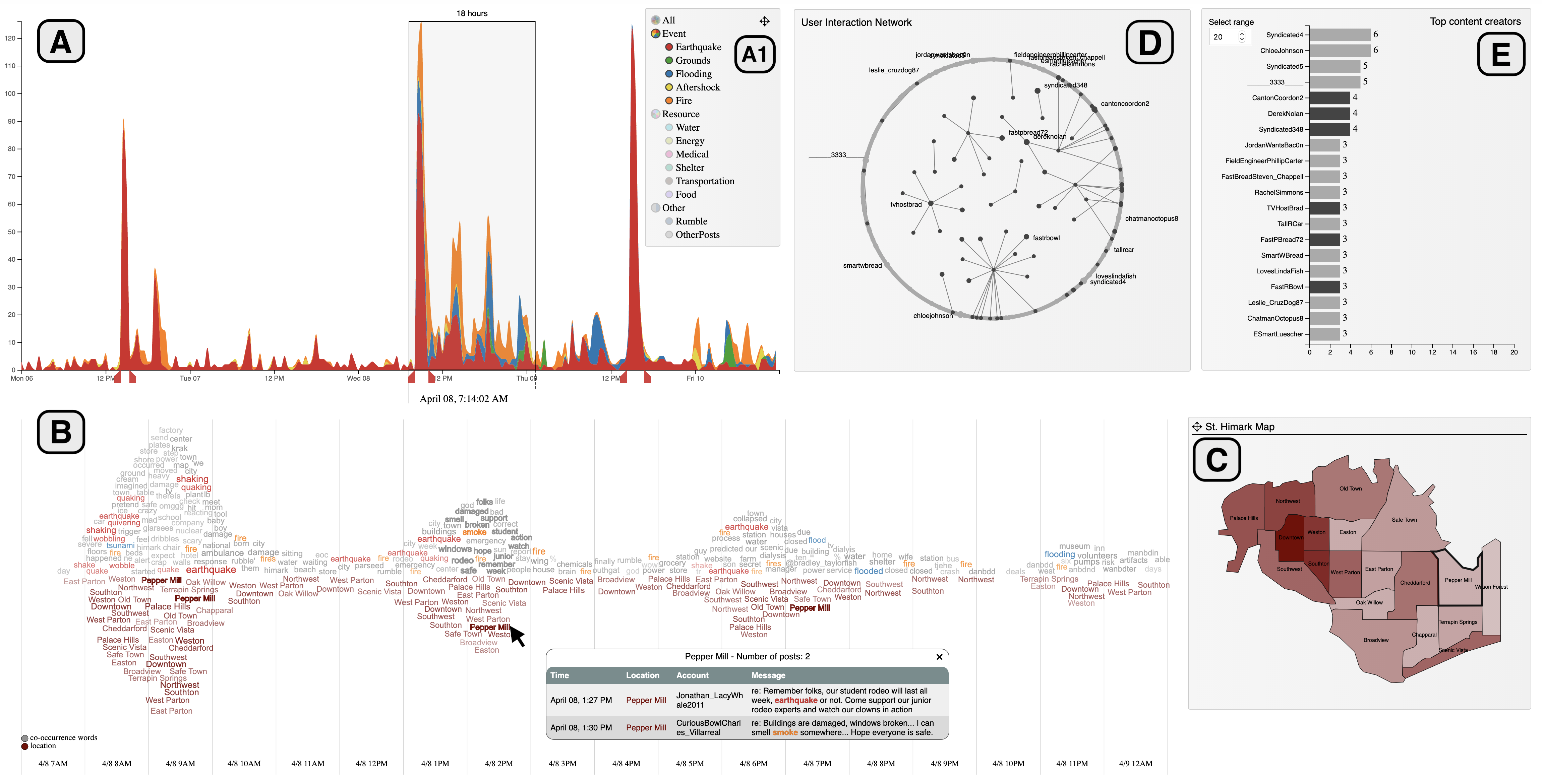}
  \caption{\label{Figure1} Application of \thename~on the \textit{VAST Challenge 2019: Mini-Challenge 3} dataset. The main control panel (A) is built as a stacked area chart, showing the volume of messages according to chosen categories from selection panel (A1). For the chosen time frame and the chosen categories: Panel (B) presents topic evolution, panel (C) is a map in which each neighborhood's color indicates the number of posts, panel (D) is a network user interaction and panel (E) is a chart for ranking content creators. Detail-on-demand and interactive features are also added to support the analysis such as sliding window with adjustable window size for updating the system and specific messages on mouse-over events (at the black arrow). In panel (A), three high red peaks correspond three times the earthquake strikes.}
  
}
\abstract{
This paper introduces \thename{}, an interactive exploratory tool for earthquake situational analytics using social media. \thename{} is designed to support users to characterize the condition across the area around the earthquake zone, regarding related events, resources to be allocated, and responses from the community. On the general level, changes in the volume of messages from chosen categories are presented, assisting users in conveying a general idea of the condition. More in-depth analysis is provided with topic evolution, community visualization, and location representation. \thename{} is developed with intuitive, interactive features and multiple linked views, visualizing social media data, and supporting users to gain a comprehensive insight into the situation. In this paper, we present the application of \thename{} with the \textit{VAST Challenge 2019: Mini-Challenge 3 (MC3)} dataset.

} 

\CCScatlist{
  \CCScatTwelve{Earthquake Analysis}{Social Media Data}{}{};
  \CCScatTwelve{Interactive Visualization}{Visual Analytics}{EQSA}{WordStream}{};
  \CCScatTwelve{VAST Challenge 2019}{}{Mini-Challenge 3}{}{}
}

\begin{document}


\firstsection{Introduction}
\maketitle
The earthquake analytics is implemented based on the general workflow including 1) characterize the overall conditions by events occurred and resources needed, then 2) discover and explore the story behind these situations on finer granularities of the data. In particular, the user utilizes the stream graph to detect the time points with a high volume of messages (for events or resources, or both), the scope of view can be adjusted by customizable time frame with a sliding window. The interactive features in supporting panels characterize the situation with the chosen range, presenting detail information and assisting users gaining insights into the dataset.

\section{Design decisions}
With a large dataset of messages from a social media platform, we first build a taxonomy for classifying these messages. For this particular earthquake dataset, we develop two main categories, which are ``event'' and ``resource''. Sub-categories for ``event'' are: earthquake, ground damage, flooding, aftershock, and fire; sub-categories for ``resources'' are: water, energy, medical, shelter, transportation, and food. Each sub-category contains a set of keywords to determine if a message belong to it or not. One message can belong to more than one category, e.g., a message can indicate that there are needs for both water and food. The reasoning for this assumption comes from the possibility that the meaning of a message can be ambiguous, hence we need to take all data into account. 

For time-series data, the stacked area chart is frequently utilized for visualizing changes of data over time, for multiple categories. We utilize this chart type to represent the volume of classified messages and provide an overall view of the condition. To explore detail, a sliding window on top of the area chart is provided for brushing.

In terms of exploring specific topics to characterize the condition, there are many approaches to handle text data. We utilize \textit{WordStream} \cite{wordstream}, a technique for visualizing evolution of topic over time. Besides the overall or global trending of terms, this technique represents the local evolution of an individual topic. Similar approach could also be found from \textit{IoTNegViz} \cite{iotnegviz}. Regarding community, a network of users represents the existing clusters, based on the relationships among users: the source is the sender and target(s) are the users that are mentioned in the message. This strategy is used to detect who are the users that have multiple connections in the community; hence, they are likely to spread the important messages. The location and ranking of users by amount of content are provided with a map and a horizontal bar chart.

\section{System model}
Figure \ref{Figure1} shows the main components of our system designed for the \textit{VAST Challenge 2019: Mini-Challenge 3} dataset\footnote{\url{https://vast-challenge.github.io/2019/MC3.html}} including a stacked area chart, a WordStream, a geographical map, a network of users and a horizontal bar chart. Detail-on-demand is provided with tooltip and applied for the WordStream, network, and bar chart.

The analytics application includes active and interactive components that can work with dynamic data. The main control panel (A) is built as a stacked area chart, showing the volume of messages according to chosen categories from selection panel (A1). The vertical axis shows the number of posts, while the horizontal axis shows the timeline. A sliding window across the timeline has expandable width from 1 hour to 31 hours. For each change in main panel (A) – whether it is adjusting time frame, timestamp or category, all other four panels (B to E) are updated according to (A). 

Panel (B) provides a WordStream, demonstrating the evolution of topics extracted from the messages' content. The WordStream consists of two topics: the keywords within the content of messages and the location of the message. The thickness of the stream is proportional to the number of posts – the global trend. Users can also explore the local trend of an individual term and detail of messages. 

Panel (C) is a geographical map, in which the color of each neighborhood indicates the number of posts. User can use this map for highlighting corresponding messages from a location in the WordStream and vice versa. 

Panel (D) is a network of user interaction. The network demonstrates the connection between users, through the account mentioning in the content of messages. Via this network, we can spot which one is the account that has an essential role in the community. 

Panel (E) is an account list for ranking content creators. This chart shows the accounts that create the largest number of posts. Users can see who writes many posts and their connection to the community; exploring these points can help detect irrelevant accounts.

\section{Implementation}
The demonstration link, explanatory video of \thename{}, and report for the challenge of \textit{VAST Challenge 2019: Mini-Challenge 3} can be found at our GitHub page\footnote{\url{https://idatavisualizationlab.github.io/VAST2019mc3/}}.

\section{\label{analysis}Analyzing VAST Challenge 2019: MC3 dataset}
\begin{figure}[t]
 \centering
 \includegraphics[width=0.99\linewidth]{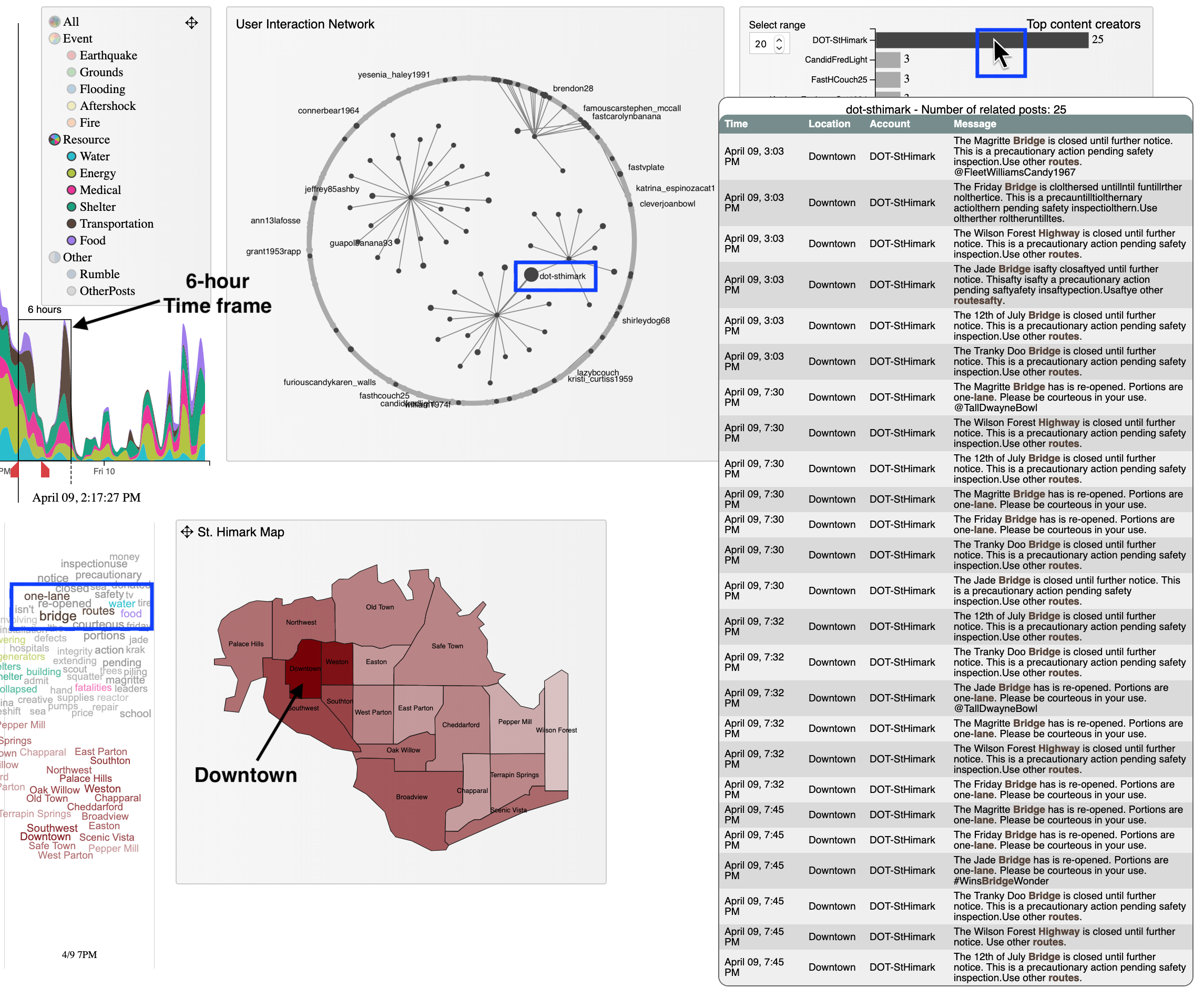}
 \caption{Five hours after the third strike, the main problem comes from ``bridges'' and ``route'' (terms emphasized in the WordStream), with multiple notifications from Department of Transportation of St. Himark.}
 \label{Figure2}
\end{figure}

Figure \ref{Figure2} shows the visualization with data from five hours after the third strike. The sliding window in the stacked area chart has a width of 6 hours. Inside the window, the brown stream - transportation has two major regions, one arise from right when the earthquake strikes and one towards the end of the window, showing that issues with transportation occurred within this time, we can come up with the observation: The problem happens and then it is addressed.


Three blue boxes are highlighted, from 1) The WordStream, 2) The network and 3) The ranking of content creators. In the WordStream, ``bridge'', ``one-lane'' and ``routes'' are emphasized, support the observation from the main panel. Moving on to the network, the node of ``dot-sthimark'', the DoT - Department of transportation of St. Himark, is in large size - an important component of the community at this time. This means that the community is connecting to the department for information. On the ranking bar chart, the department provides 25 notifications within 5 hours, a large number comparing to the next ranked account (only 3). In the map, the neighborhood of Downtown is in the darkest shade - with biggest number of posts, because the DoT is located in Downtown. By exploring the detail messages from the department, we discover that the bridges to get in and out of St.Himark are closed and then re-opened for safety inspection. This confirmation helps validate the observation mentioned above.



\section{Conclusions and future works}
This paper proposes an interactive exploratory visualization to analyze earthquake and characterize the condition from social media posts. The general view and brushing/linking features support users to explore and discover the patterns and relating events. The application of this technique to \textit{VAST Challenge 2019: Mini-Challenge 3} dataset also confirms the design decisions of multiple linked views help users to convey the messages from the perspectives of community, location and volume of related posts. Future work will focus on optimization of the overall layout and scalability to big data.

\bibliographystyle{abbrv-doi}

\bibliography{template}
\end{document}